# EXPERIENCES WITH ADVANCED CORBA SERVICES

G. Milcinski, M. Plesko, M. Sekoranja, Josef Stefan Institute, Ljubljana, Slovenia

Abstract

The Common Object Request Broker Architecture (CORBA) is successfully used in many control systems (CS) for data transfer and device modeling. Communication rates below 1 millisecond, high reliability, scalability, language independence and other features make it very attractive. For common types of applications like error logging, alarm messaging or slow monitoring, one can benefit from standard CORBA services that are implemented by third parties and save tremendous amount of developing time. We have started using few CORBA services on our previous CORBA-based control system for the light source ANKA [1] and use now several CORBA services for the ALMA Common Software (ACS) [2], the core of the control system of the Atacama Large Millimeter Array. Our experiences with the interface repository (IFR), the implementation repository, the naming service, the property service, telecom log service and the notify service from different vendors are presented. Performance and scalability benchmarks have been performed.

## 1 INTRODUCTION

Our team has over the last five years developed a control system framework that uses and extends modern component-based, distributed computing and object-oriented concepts. The basic entities of the system are accelerator devices that are represented as CORBA (Common Object Request Broker Architecture) objects – objects that are remotely accessible from any computer through the established client-server paradigm. We chose CORBA among other environment for distributed systems (CDEV, J2EE, DCOM...) because of its platform and language independence. A successful implementation, based on Borland's Visibroker [6] is running the CS at the light source ANKA. It uses Visibroker's proprietary smart agent, location service and interface repository. Then we decided also to use other standard CORBA services. At first we had some doubts – we were not sure if we could accommodate to programs that were not written by us, we were afraid of the high prices of some of these programs, etc. But starting fears have vanished quickly. We have completely rewritten the framework, making it more general and useful for other control systems [2]. We used mostly TAO [4] and ORBacus [5] implementations (both are free for non-commercial use), both for ORB and services.

## 2 USED SERVICES

### 2.1 Event Service

This service coordinates the communication between two kinds of objects – supplier (it produces event data) and consumer (it processes event data). That is exactly what a control system is doing – for example: user sets current to a power supply and vice versa – a machine sends readback to the user. When we started to develop ANKA control system, there was a major disadvantage of this service – it supported just generic events of type Any. But, to discover all typing errors at compile time, we wanted typed events [1]. That is why we defined our own callback classes, one for each data type.
Nowadays both ORBacus and TAO Event Service already support typed events (but TAO's solution for typed events support is non-standard).

### 2.2 Notify service

It extends Event Service and has added some further functionality. These are filtering events (by type and data), subscription to only some kinds of events, the ability to configure various qualities of service properties (per-channel, per-proxy or even per-event).
This is in our opinion the most useful service. It is just perfect for controlling a few devices. The main problem is, that queuing, filtering and processing events demand time, memory and CPU and it could not process all data used in a large control system It is a potential bottleneck and a single point of failure. It is best used for system wide services such as alarm and logging, where one central process collects all messages from anywhere in the control system.
ORBacus and TAO Notify Service are not supporting typed events. One can use structured events, which are actually Anys, but you can set a type property of an event.
We performed benchmark and scalability tests on this service, which will be discussed later.

### 2.3 Telecom Log Service

It is some kind of event consumer, which stores data in persistent store. In some cases it must also supply an event (to inform user that something in its state has changed – like when a threshold is being crossed). ORBacus did actually implement notify logging service which has all notification functionality and a persistent store. User can also query log entries, using some kind of filter.

ORBacus T-Log has already implemented storing data in its own database, which is not suited for large amounts of data. TAO's Telecom Log Service stores records in memory and it is actually just a skeleton for a serious implementation. We had to add features ourselves (like persistent store, sending events to notify channel, etc).

## 2.4 Naming Service

A useful tool - just like the telephone directory. It is used to give names to objects. To work properly there are two requirements – all objects have to be named and each name is used only once. An object can have two names, but vice versa is not possible. It is much like file structure on hard disk – the equivalent name can only be used for a file in different directory. Other services are easer to manage when connected to naming service (for example: notify supplier and consumer can exchange IOR-s of the event channel; in ORBacus demos you can find an approach without NS – saving and reading IOR to/from a file – a little clumsy idea - just think about sending a file by every channel creation).

## 2.5 Property Service

Property Service introduces a Property Set, which is a collection property. Every property has a name (unique within the property set) and a value, which can be of any type (the CORBA *any* type). Property Sets are very useful for storing object's data. For example, an object representing a device in our control system is storing all its characteristics in a property set, so that they are all read in a single CORBA call during initialization. On Windows systems, a key in the registry is equivalent to a property set.

## 2.6 Interface Repository

A service that exposes the interfaces of CORBA objects (the IDL file) in form of an object model, which is available at run-time. Through the IFR, a program is able to determine what operations are valid on an object and make an invocation on it, even though its interface was not known at compile-time In that way we have developed Object Explorer (OE) – a program, which can control the whole system without knowing almost anything about the structure of the controlled devices. The OE finds all CORBA objects on the network and asks IFR for their operations. Using dynamic invocation interface (DII) it executes chosen method via its name and queries the user for all parameters in the parameter list.

Another interesting usage of the IFR was in our JavaBeans generator: the generator queries the running IFR and creates Java source code that wraps the CORBA objects into Java Beans – one Bean per CORBA object interface. This is much more convenient that writing our own IDL parser. We use ORBacus implementation of this service and have found it very stable – it has been running continuously for three months now.

## 2.7 Implementation Repository

The Implementation Repository contains information that allows the ORB to locate and activate implementations of objects. Ordinarily, installation of implementations and control of policies related to the activation and execution of object implementations is done through operations on this service.

We did not actually use this service, but took its features and interfaces into account when writing our main management program. It starts objects, loads their shared libraries and other CORBA services, needed for logging, archiving, etc.

# 3 PERFORMANCE AND SCALABILITY BENCHMARKS

Tests were made with 1 GHz Athlon PC with 512 Mb of RAM. All processes were running on the same machine, so network latencies were excluded. The downside of this is that processes switching might have affected the results. We can safely assume that the real performance is only better. All test were made with notify service's default settings.

We concentrated on testing a Notify Service for three reasons. It is the easiest to test, results represent also event service performance and it is the only one whose performance directly influences the performance of the CS Already on the beginning we found a minor advantage of TAO notify service – when it starts, it writes an IOR of an event channel factory to the name service. ORBacus notify does not, so user must do it manually or use resolve_initial_reference instead.

## 3.1 Time needed for processing an event

First test is very simple. We have one supplier and one consumer. The supplier sends events and the consumer receives them, both doing it as quickly as they can. Trying to overload the service, supplier was sending events in separate threads. First observation in the figure 1 is that time, needed for one event, is increasing with number of threads (except from one thread to ten threads – this is expected because of better exploiting of CPU). A big jump from 30 threads to 100 threads is presumably consequence of overloading the CPU.

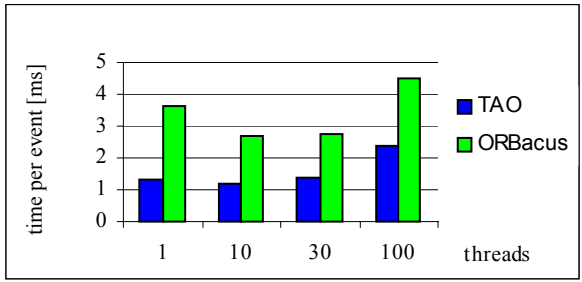

**Figure 1 - chart shows average time, needed for processing an event (processes for sending events were running on separate threads)**

We can also see that TAO is quite faster than ORBacus. In this test we have also noticed TAO's immunity to increasing number of events. This cannot be said for ORBacus, which had quite a few problems dealing with this. It actually lost a bit of them, which can be very critical in some conditions.

### 3.2 Increasing number of suppliers

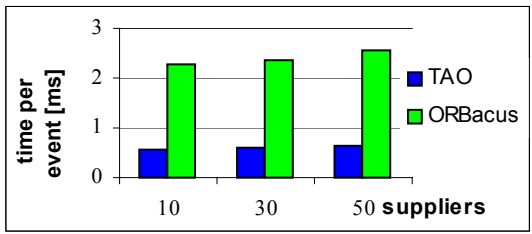

**Figure 2 - average time for processing an event from many suppliers**

From 10 to 50 suppliers were connected to the same event channel and doing their job. The result is quite expected. Time needed for one event is slowly increasing (from 0.5 ms, 10 suppliers, to 0.6 ms, 50 suppliers at TAO and 2.3 to 2.6 ms, ORBacus). We can again see that TAO is faster.

### 3.3 Increasing number of consumers

We used one supplier, to which 10 to 50 consumers were connected.
The time per event increases with the number of consumers, because the service must create one event for each consumer. Although ORBacus is slower again it has one advantage. If we divide time needed for one event with number of consumers, we get the result, which can be seen in figure 2.
With ORBacus, the needed time is decreasing (just the opposite from TAO). This means that ORBacus is better dealing with big number of consumers. This can be probably explained with better logic for multiplying events.

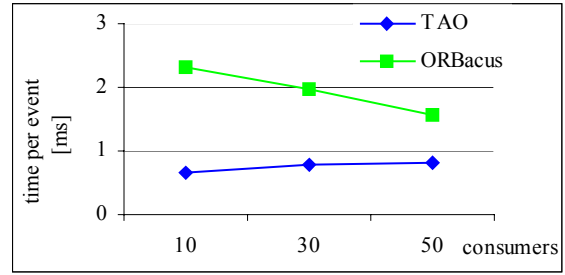

**Figure 3 - chart shows average time needed for multiplying an event - one supplier sends event, many consumers are receiving it**

### 3.4 Other observations

First thing you notice dealing with these two services is much more professional appearance of ORBacus. Web pages are clearer, documentation is extensive, replies to messages, send to mail list, are quicker. ORBacus service is also more thoroughly implemented (more possibilities for quality of service settings, etc). It is also easier to destroy TAO service with a bad client. These advantages are probably at the same time the reasons, which make ORBacus service slower.

## 4 CONCLUSIONS

Most of the services, described in this paper, are successfully running at ALMA [2] (for now just for testing and development purposes), ANKA [1] and other systems. Many others and ourselves have found them very useful – so it is a waste of time and money not to use them. But they are just not so perfect to use them without fundamental consideration of possible bottlenecks and other points of failure. And we still hat to implement a few extra features of our own, so it is unlikely to get away without programming..